\documentclass[11pt]{article}
\usepackage[english]{babel}
\usepackage{amsmath}
\usepackage{amsthm}
\usepackage{amssymb}
\usepackage{graphicx,epstopdf,subfigure,color,hyperref,enumerate}
\newcommand{\beq}{\begin{equation}}
\newcommand{\eeq}{\end{equation}}

\newcommand{\ltwid}{\mathrel{\raise.3ex\hbox{$<$\kern-.75em\lower1ex\hbox{$\sim$}}}}
\newcommand{\gtwid}{\mathrel{\raise.3ex\hbox{$>$\kern-.75em\lower1ex\hbox{$\sim$}}}}
\input epsf

\usepackage{graphicx}

\title{\bf What weak measurements and weak values really mean - Reply to Kastner}

\author{E. Cohen\footnote{H.H. Wills Physics Laboratory, University of Bristol, Tyndall Avenue, Bristol, BS8 1TL, U.K}}

\begin{document}
\maketitle

\begin{abstract}
Despite their important applications in metrology and in spite of numerous experimental demonstrations, weak measurements are still confusing for part of the community. This sometimes leads to unjustified criticism. Recent papers have experimentally clarified the meaning and practical significance of weak measurements, yet in [R.E. Kastner, Found. Phys. 47, 697–707 (2017)], Kastner seems to take us many years backwards in the debate, casting doubt on the very term ``weak value'' and the meaning of weak measurements. Kastner appears to ignore both the basics and frontiers of weak measurements and misinterprets the weak measurement process and its outcomes. In addition, she accuses the authors of [Y. Aharonov et al., Ann. Phys. 355, 258-268 (2015)] in statements completely opposite to the ones they have actually made. There are many points of disagreement between Kastner and us, but in this short reply I will leave aside the ontology (which is indeed interpretational and far more complex than that described by Kastner) and focus mainly on the injustice in her criticism. I shall add some general comments regarding the broader theory of weak measurements and the Two-State-Vector Formalism (TSVF), as well as supporting experimental results. Finally, I will point out some recent promising results, which can be proven by (strong) projective measurements, without the need of employing weak measurements.
\end{abstract}
\vskip 1cm

Weak measurements and weak values \cite{AAV,Unified,Coll} are attracting an increasing amount of attention thanks to their successful assistance in solving conceptual \cite{Slits,Past,Genovese1,Genovese2} and practical questions \cite{Tech1,Tech2,Tech3,Tech4,Tech5,Tech6}. Recent experimental evidences strengthen the profound quantum nature of weak values \cite{NewStein} and their physical meaning, which goes far beyond a conditional average \cite{NewVaid}. 

In what follows, I examine and disprove several of Kastner's claims \cite{REK}, but first I shall correct some injustice made to the authors of \cite{ACE}, which Kastner denotes by ACE. She asserts that ``It should also be clarified that taking postselection into account does not indicate any departure from standard one-vector quantum theory, as ACE suggest.'' In fact, we have never suggested that. This claim of Kastner stands in stark contrast with the statements  that Aharonov, Elitzur and I have made in \cite{ACE}: ``As TSVF and traditional quantum theory are equivalent, obliging one- and two-vector explanations to be equally valid, this contradiction can be resolved in two ways'' and ``TSVF is unique among the above models in that it has derived several predictions that, although fully consistent with the standard formalism (see Appendix B), seem surprising and more acutely
opposed to classical laws''. Moreover, we have explained in Appendix B of \cite{ACE} that the resulting probabilities in the one- and two-state-vector approach are identical. Kastner's repetitive claims that weak measurements and weak values are part of standard quantum mechanics, are therefore obvious and well-known. To suggest that ACE think differently is highly misleading, as we obviously claimed otherwise.

First, we have argued that weak values simplify calculations. In \cite{ACE}, for instance, one had to calculate the measurement outcomes of 9 sequential weak measurements. This could be a daunting task when one considers a standard forward-in-time calculation followed by projection on the final state. But when using weak values, and thus both pre- and postselection at the same time, all outcomes can be easily found (up to minor corrections which scale like the square of coupling strength).

Our second claim was that weak values can be insightful and this is actually an historical fact. In addition to the above mentioned practical importance, weak measurements have led to the discovery of superoscillations \cite{SO1,SO2}; to the development of quantum random walks \cite{QRW}; and to a fresh look on many paradoxes \cite{AR} such as Hardy's \cite{Hardy,HardyW}, to name just a few examples.

Regarding the definition of weak values, Kastner states ``It is important to note that this is a theoretical quantity defined in terms of operators and
states, without regard to any particular process of measurement. Thus, the term `weak value' is something of a misnomer: there is nothing `weak' about the value itself.'' This claim can be mathematically falsified, as the weak value naturally emerges when employing the linear approximation of a system's time evolution. This approximation is usually possible when the coupling is weak enough, and upon using the von Neumann measurement scheme. Incidently, the well-known derivation below has recently appeared in a book which Kastner co-edited \cite{CA}.

Let us, then, employ the following von Neuamnn interaction Hamiltonian

\begin{equation}
H=H_{\textrm{int}}=g(t)A\otimes p_{d},
\end{equation}
where $A$ is the observable to be measured, $p_d$ is the momentum of the pointer (canonically conjugated to its position $q$) and $\int_{0}^{T}g(t)dt=g$ for a coupling time $T$ and coupling strength $g$.

When performing a weak measurement on a pre- and postselected ensemble $\langle \phi|~|\psi \rangle$ using a measurement pointer with initial wave function described by $|\Phi(q)\rangle$, the time evolution of the measured system + pointer is:

\begin{equation}
\begin{array}{lcl}
\langle \phi| e^{-i\int H dt/\hbar}|\psi\rangle\otimes|\Phi(q)\rangle \approx
\langle \phi| 1 -ig A\otimes p_{d}|\psi\rangle\otimes|\Phi(q)\rangle =
\langle \phi|\psi\rangle \left(1-ig\langle A \rangle_w p_{d}\right)|\Phi(q)\rangle \approx \\
\langle \phi|\psi\rangle e^{-ig\langle A \rangle_w p_{d}}|\Phi(q)\rangle =
\langle \phi|\psi\rangle |\Phi(q-g\langle A \rangle_w)\rangle
\end{array}
\end{equation}

Hence, the weak value defined by

\begin{equation} \label{ECWV}
\langle A \rangle_w= \frac{\langle \phi |A| \psi \rangle}{\langle \phi | \psi \rangle}
\end{equation}\\

naturally emerges. Yes, weak values can be seen as normalized transition amplitudes as Kastner insists, but it does not mean they do not possess the above important property. One then may choose to ignore the physics leading to the emergence of weak values (for instance, the reason we need to normalize these transition amplitudes), as well as their general appearance when two systems are weakly coupled \cite{Potential}. But then one may reach incomplete conclusions.

Furthermore, Kastner then ventures to assert that weak values merely reflect a statistical quantity, rather than a quantity pertaining to each individual system in the measured ensemble. She further compares weak values to classical averages in this respect. This, however, was recently challenged on experimental grounds by \cite{Genovese1,Genovese2} and \cite{NewStein} showing that weak values reveal themselves in the {\it single particle level}, and moreover by \cite{NewVaid}, proving that a weak value resembles more an eigenvalue than a conditional expectation value. It is worth mentioning that in all the above experiments, weak values exhibited {\it inherently quantum} features, hence the comparison to a classical averages is misleading.

I now turn to what seems to be Kastner's main argument, namely, ``Assertions in the literature that weak measurements leave a system negligibly disturbed... are therefore unsupportable''. In light of the above, this claim might seem shocking, but it rests on three basic misunderstandings:\\
1. Kastner writes: ``Clearly, therefore (unless	we have exactly $a=b=1/\sqrt{2}$) $S$ has been non-negligibly disturbed''. This, in my opinion, reflects a misunderstanding of the term ``negligibly''. In the case described by Kastner, the system is absolutely not affected by the measurement, so the adverb ``negligibly'' is not in place. If, however, $a=[(1+\epsilon)/2]^{1/2}$ and $b=[(1-\epsilon)/2]^{1/2}$, for $\epsilon \ll 1$, which is indeed justified in the weak measurement regime, then the fidelity is 1 up to $O(\epsilon^{2})$. Therefore, by definition, the state has been negligibly disturbed, and this is now a precise claim. In fact, it is this unique regime granting weak measurements their significance.\\
2. An ancilla-based measurement, like the ones ACE and Kastner analyze, is commonly used. Clearly, during the measurement process entanglement between the measured system and the measuring pointer is created and then vanishes, but this fact alone does not tell us how much the measured state has changed. This disturbance depends on the coupling strength between the systems. When performing a projective measurement, the systems are strongly entangled, and hence a change in the pointer state definitely reveals itself in the state of the measured system. The power of weak measurements lies in the weak entanglement between the system and pointer, which upon a strong measurement of the pointer breaks down, but leaves only a minute change in the system's state. Therefore claims such as ``The only difference between this `weak	measurement' and the standard `strong' or `sharp' measurement is that $S$ is not in	an eigenstate of	its	observable'' are erroneous and miss the importance of weak measurements.\\
3. As I emphasized in \cite{EConFC}, weak measurements are non-invasive, that is, the probability of evolving the initial state to an orthogonal state through a weak measurement decreases like $g^2$. This property significantly limits the amount of backaction, but seems to be ignored by Kastner.

Kastner also suggests that: ``ontological claims based on such assertions need to be critically reassessed''. The problem with the last sentence is that Kastner seems to be unfamiliar with recent ontological claims \cite{CA,Collapse,Entropy,Ont,Heis}. It is unfair to judge an ontology without referring to the actual works presenting it. The fallacy here is obvious: The TSVF does not provide an ontology. It is the Two-Time Interpretation \cite{Gruss} further developed either within the Schr\"{o}dinger picture \cite{CA,Collapse,Entropy} or within the Heisenberg picture \cite{Heis} that outlines the ontology, but Kastner does not address these works. The claims regarding ontology are therefore completely unsupported.

Finally, it is worth pointing out a growing number of recent promising results which enable clarifying the TSVF predictions using {\it strong, non-counterfactual measurements} \cite{OT,ACLE,Liar}. Being non-statistical in character, they are therefore immune to this criticism of Kastner, as well as to any criticism of the TSVF based on weak measurements. Moreover, they accord well with the school to which Kastner herself adheres, namely, the Transactional Interpretation. In some sense, these advances agree with Kastner's demand that weak values should be revealed by sharp measurements, but we refer only to special cases where weak values coincide with the measured projector's eigenvalues. Even in light of those cases we remember that weak measurements and weak values have provided throughout the years an invaluable source of both helpful tools and insightful clues for addressing foundational topics in quantum theory.\\ 

\noindent{\bf Acknowledgements}\\
This work was supported by ERC AdG NLST.

\end{document}